\documentclass[10pt,twocolumn]{revtex4}
\usepackage{graphicx,subfigure,epsfig,appendix}
\usepackage{booktabs,multirow,mathrsfs,amssymb,amsmath,amsthm}
\usepackage{algorithm,algorithmicx,framed,listings} 
\usepackage[colorlinks=true,linkcolor=blue,citecolor=blue,urlcolor=blue]{hyperref}
\usepackage[noend]{algpseudocode}
\usepackage{pifont}
\newcommand{\bra}[1]{\langle #1|}
\newcommand{\ket}[1]{|#1\rangle}

\newcommand{\ds}[2]{\mathbb{DS}(#1,#2)}
\graphicspath{{figures/}}
\begin{document}

 
\title{Enhancing the expressivity of quantum neural networks with residual connections}

\author{Jingwei Wen}
\email{wjw17@tsinghua.org.cn}
\author{Zhiguo Huang}
\author{Dunbo Cai}
\author{Ling Qian}
\email{qianling@cmss.chinamobile.com}

\affiliation{China Mobile (Suzhou) Software Technology Company Limited, Suzhou 215163, China}

 
\begin{abstract}

In the recent noisy intermediate-scale quantum era, the research on the combination of artificial intelligence and quantum computing has been greatly developed. Inspired by neural networks, developing quantum neural networks with specific structures is one of the most promising directions for improving network performance. In this work, we propose a quantum circuit-based algorithm to implement quantum residual neural networks (QResNets), where the residual connection channels are constructed by introducing auxiliary qubits to the data-encoding and trainable blocks of the quantum neural networks. Importantly, we prove that when this particular network architecture is applied to a $l$-layer data-encoding, the number of frequency generation forms can be extended from one, namely the difference of the sum of generator eigenvalues, to $\mathcal{O}(l^2)$. And the flexibility in adjusting the corresponding Fourier coefficients can also be improved due to the diversity of spectrum construction methods and the additional optimization degrees of freedom in the generalized residual operators. These results indicate that the residual encoding scheme can achieve better spectral richness and enhance the expressivity of various parameterized quantum circuits. Extensive numerical demonstrations in regression tasks of fitting various functions and applications in image classification with MNIST datasets are offered to present the expressivity enhancement. Our work lays the foundation for a complete quantum implementation of the classical residual neural networks and explores a new strategy for quantum feature map in quantum machine learning.

\end{abstract}

\maketitle

\section{introduction}

Quantum computing is a new computing paradigm based on quantum mechanics that utilizes qubits instead of classical bits to store and process information \cite{nielsen2010quantum}. Since the theoretical concepts were proposed \cite{feynman1982simulating,benioff1980computer,deutsch1985quantum}, quantum computers have developed at an astonishing speed, gradually moving from the proof-of-principle demonstration like quantum supremacy in the laboratory \cite{arute2019quantum,zhong2020quantum,wu2021strong} to the stage of application exploration \cite{cao2019quantum,cumming2022using,herman2022survey}. Among its many applications, quantum machine learning is an emerging field that leverages the power of quantum computers to overcome bottlenecks of high computing power requirements in the machine learning \cite{schuld2015introduction,biamonte2017quantum,cerezo2022challenges,zeguendry2023quantum}. On the current noisy intermediate-scale quantum devices \cite{preskill2018quantum}, one popular strategy for constructing quantum machine learning algorithms is using classical-quantum hybrid optimization loops to train the parameterized quantum circuits for various learning tasks, such as pattern recognition \cite{li2020quantum,henderson2020quanvolutional} and classification \cite{havlivcek2019supervised,farhi2018classification,hur2022quantum}. 

Similar to the classical neural networks that consist of input layers, hidden layers and output layers, the fundamental structures of the variational quantum neural networks include data-encoding or quantum feature map circuits $\mathcal{U}(x)$, which map the classical data $x\in\chi$ to a quantum state in Hilbert space $\mathcal{H}$, variational ansatz $\mathcal{W}(\theta)$ containing trainable parameters $\theta$, and output layers realized by quantum measurement \cite{beer2020training,abbas2021power}. To be specific, the data-encoding processes serve as one of the main sources of non-linearity for the networks, and there exist numerous encoding strategies such as amplitude encoding and angle encoding \cite{schuld2019quantum}. Moreover, different choices of architectures for the variational ansatz will lead to various quantum neural networks \cite{dallaire2018quantum,cong2019quantum,chalumuri2021hybrid,wu2021application,wang2021quantum,landman2022quantum,bausch2020recurrent,liu2022quantum,kashif2023resqnets} and it will greatly affect the network performance such as generalization \cite{banchi2021generalization} and trainability \cite{mcclean2018barren}. For example, general deep parameterized quantum circuits suffer from the barren plateau phenomenon, leading to vanishing gradients \cite{mcclean2018barren,cerezo2021cost,marrero2021entanglement,wang2021noise}. But it can be avoided by networks with hierarchical structure, proposed as a realization of the quantum convolutional neural networks (QCNN) \cite{cong2019quantum,herrmann2022realizing,hur2022quantum}, which has been proved the absence of barren plateaus \cite{pesah2021absence}. Finally, the output of an $n$-qubit quantum neural networks is the mean value of a measurable observable $O$ as
\begin{equation}
f(x,\theta)=\bra{\psi_{0}}U_{\theta}^{\dagger}(x)OU_{\theta}(x)\ket{\psi_{0}}
\end{equation} 
where initial state $\ket{\psi_{0}}=\ket{0}^{\otimes n}$ and $U_{\theta}(x)$ is the parameterized quantum circuit consisting repeatable data-encoding and trainable blocks. Interestingly, the expressivity and universality of such variational quantum models can be guaranteed by the fact that one can naturally write the outputs as partial Fourier series in the network inputs \cite{gil2020input,perez2020data,caro2021encoding,schuld2021effect}, and the accessible frequencies are determined by the eigenvalues of the generator Hamiltonian in the data-encoding gates, while the coefficients are controlled by the design of the entire circuits \cite{schuld2021effect}. 

\begin{figure*}
\centering
\includegraphics[width=\linewidth]{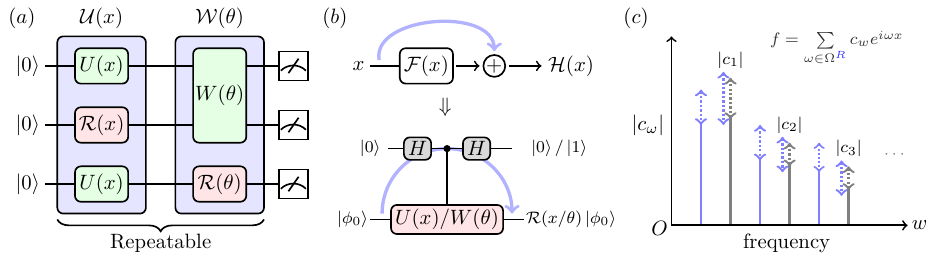}
\caption{(a) A schematic of the quantum neural networks with residual connections. The quantum feature map circuit $\mathcal{U}(x)$ and trainable variational circuit $\mathcal{W}(\theta)$ are repetitively implemented multiple times to form the multilayer structures. The $\mathcal{R}(x/\theta)$ blocks labeled by red represent the data-encoding gates $U(x)$ and parameterized gates $W(\theta)$ with residual connections. (b) The classical residual unit and its quantum counterpart. The residual connection channels are shown with blue arrows, and the output of residual block is $\mathcal{H}(x)=\mathcal{F}(x)+x$, where non-linear function $\mathcal{F}(x)$ represents the classical neural networks. The quantum residual operator $\mathcal{R}(x/\theta)$ implemented on the initial state $\ket{\phi_{0}}$ can be realized in the subspace of an ancillary qubit with measurement results $m_a=0/1$. (c) The residual feature map can introduce more frequency components (blue) to the original spectra of quantum neural networks (gray), and also make the Fourier expansion coeﬀicients more flexible.}
\label{fig-flow}
\end{figure*}

A great deal of research work has subsequently devoted to advancing the quantum neural networks, with one intuitive approach being the quantization of classical networks \cite{landman2022quantum,bausch2020recurrent,liu2022quantum,kashif2023resqnets}. Especially, inspired by the classical residual neural networks, which are proposed for alleviating the vanishing gradient problem during the training process of deep neural networks \cite{he2016deep}, its quantum counterpart is promising to mitigating barren plateaus \cite{kashif2023resqnets}. The key idea is to introduce residual connections into the traditional neural networks, as shown in the figure \ref{fig-flow}. Mathematically, the residual connections can provide an additional cross-layer propagation channel for the input features, leading to a basic residual unit of neural networks as $\mathcal{H}(x)=\mathcal{F}(x)+x$, where the non-linear parameterized function $\mathcal{F}(x)$ represents the traditional neural networks. Although there exist some works on the quantum realization of residual neural networks, the residual channels are usually implemented using classical or hybrid methods \cite{kashif2023resqnets,shi2023hybrid}. The researches on the full quantum implementations of residual connections and effects on the expressivity are still very lacking.

In this work, we address these issues by proposing a quantum algorithm for the digital simulation of quantum residual neural networks (QResNets). The residual connection channel is constructed through one ancillary qubit and the target evolution process is embedded in the subspace. Such structures are compatible to both the data-encoding and trainable blocks in the variational quantum neural networks. We also further parameterize the encoding gates on the auxiliary qubit and obtain the generalized residual operators. Furthermore, we find that the Fourier spectrum of the output of parameterized quantum circuits can be enriched when the residual connections are used for the data-encoding blocks. The number of frequency combinations forms can be extended from one, namely the difference of the sum of generator eigenvalues, to $\mathcal{O}(l^2)$ for the $l$-layer residual encoding. Moreover, the diverse construction methods for frequencies in the residual loss functions and the extra trainable parameters in the generalized residual operators can expand the Fourier coefficient space. The results suggest that the expressivity of quantum models can be enhanced by residual connections. We offer extensive numerical demonstrations of the quantum algorithm in the regression tasks by function fitting of Fourier series, and also present the performance of binary classification with standard MNIST datasets to recognize the handwritten digits images, achieving an accuracy improvement of over 7\% with residual encoding.

The remainder of this paper is organized as follows. We introduce the theory in the Sec \ref{ChapTheory}, including realization of quantum residual connections, proof of frequency spectra enhancement and measurement scheme. Sec \ref{ChapSim} and Sec \ref{ChapExp} give the numerical results of the proposed quantum algorithms in fitting functions and classifying handwritten character images. Finally, a conclusion in the sec \ref{ChapCon} is given.

\section{theory}\label{ChapTheory}

\subsection{Realization of Quantum Residual Connection}

In the QResNets, there are multiple layers of repeatable data-encoding block $\mathcal{U}(x)$ and trainable parameterized ansatz $\mathcal{W}(\theta)$, and the residual connections can be adopted in some of the blocks, as shown in the figure \ref{fig-flow}. The data-encoding block consists of quantum rotation gates of the form $U(x)=e^{iHx}$ where $H$ is a generator Hamiltonian, while the trainable circuits are composed of single- and two-qubit parameterized quantum gates $W(\theta)$ with optimization parameters $\theta$. Some gates in the data-encoding and ansatz block can be sampled to add residual connections forming quantum residual operators $\mathcal{R}(x/\theta)$, which correspond to the residual evolution process. For an $n$-qubit quantum system with initial state $\ket{\phi_{0}}$, the evolution under residual operator can be expressed as 
\begin{equation}
\mathcal{R}(x/\theta)\ket{\phi_{0}}=\frac{1}{2}(\sigma_{0}^{\otimes n}+\mathcal{L}(x/\theta))\ket{\phi_{0}}
\label{R-resnet}
\end{equation} 
where $\sigma_{0}$ is the identity matrix and $\mathcal{L}(x/\theta)$ is a unified expression of the gates in data-encoding and trainable blocks. It means that $\mathcal{L}(x/\theta)=U(x)$ in the quantum feature map block and $\mathcal{L}(x/\theta)=W(\theta)$ in optimization ansatz. Such an evolution operator can be realized by the frame of linear combination of unitary with one ancillary qubit, and the target quantum states are obtained by post-processing \cite{gui2006general,childs2012hamiltonian}. Specifically, we first apply a Hadamard gate to encode the ancillary system followed by a controlled-$\mathcal{L}(x/\theta)$ operator. After adding another Hadamard gate, we can measure the ancillary qubit with results $m_{a}=0/1$ corresponding to quantum states $\ket{0}/\ket{1}$. Then the evolution results under residual operators can be obtained in the $\ket{0}\bra{0}$ subspace. The introduction of an auxiliary qubit provides an additional channel that allows the unevolved quantum state to pass alone and add to the evolved quantum state. 

More generally, the weight of the summation process can also be adjusted by replacing the first Hadamard gate on the ancillary qubit with $R_{y}(2\alpha)$ rotation with trainable angles $\alpha$. Then the corresponding residual operator is generalized as a single optimization-angle residual operator 
\begin{equation}
\begin{split}
\mathcal{R}_{1}(x/\theta)=\frac{\cos\alpha \sigma_{0}^{\otimes n}+(-1)^{m_a}\sin\alpha\cdot\mathcal{L}(x/\theta)}{\sqrt{2}}
\label{R1op}
\end{split}
\end{equation} 
Such a construction does not require a post-selection process, but rather reconstructs the target operator from the measurement results. It can be reduced to $\mathcal{R}(x/\theta)$ with $\alpha=\pi/4$ and $m_a=0$. Similarly, a two optimization-angles residual operator $\mathcal{R}_{2}(x/\theta)$ can also be constructed by replacing both Hadamard gates with parameterized rotation gates, and the detail is shown in appendix \ref{appendixR2}. In principle, the introduction of more trainable parameters in these two generalized residual operators will provide additional degrees of freedom for optimization, which can further increase the expressivity of the parameterized quantum circuits.

Therefore, we can conclude here that a general residual connection in quantum neural networks can be realized in the complete quantum circuit frame. It is also worth noting that in some special network structures such as the QCNN \cite{cong2019quantum}, by reusing discarded qubits, we can simulate the residual connections without additional qubits. Moreover, due to the fact that the expressivity of quantum models is fundamentally limited by the data-encoding strategy, we will prove below that the residual connections applied to data-encoding block, no matter what ansatz used, will lead to a better spectra richness in the Fourier series of quantum model output, resulting an expressivity enhancement.


\subsection{Frequency Spectra Enhancement}

It has been pointed out that the output of a parameterized quantum circuit can be expressed as a finite-term Fourier series of the input features \cite{schuld2021effect}
\begin{equation}
\begin{split}
f(x,\theta)=\sum_{\omega\in\Omega}c_{\omega}(\theta,O)e^{i\omega x}
\end{split}
\label{fourier}
\end{equation} 
where the frequency $\omega$ of spectrum $\Omega=\{w_{k}-w_{j}|j,k\in [d]\}$ depends on the $d$-dimensional generator of one-layer data-encoding gate $U(x)=e^{iHx}$ with eigenequations $H\ket{h_{j}}=w_{j}\ket{h_{j}}$ for $j\in [d]$. Notation $[d]:=\{1,2,\cdots,d\}$ here. It means that the accessible frequency of the quantum model is constructed from the difference between the generator eigenvalues. For example, a frequently used generator is the Pauli matrix $H=\sigma/2$ with two eigenvalues $w_{1,2}=\pm1/2$ where $\sigma=\{\sigma_{x},\sigma_{y},\sigma_{z}\}$, then such a one-layer data-encoding block would produce a frequency spectrum $\Omega=\{0,\pm1\}$. Moreover, the expansion coeﬀicients $c_{\omega}(\theta,O)$ are associated with the entire structure of the quantum circuit including trainable parameters $\theta$, and the observable $O$. 

However, for a data-encoding block with residual connection, more frequency components can be involved, realizing an improvement in the circuit approximation ability. Assuming that the initial quantum state $\ket{\phi_{0}}$ of  the residual encoding block is related to the optimization parameters $\theta$, the residual loss function can be expressed as 
\begin{equation}
\begin{split}
f_{R}(x,\theta)=&\bra{\phi_{0}}\mathcal{R}^{\dagger}(x)O\mathcal{R}(x)\ket{\phi_{0}}\\
=&\frac{1}{4}\big(\bra{\phi_{0}}U^{\dagger}(x)OU(x)\ket{\phi_{0}}+\bra{\phi_{0}}O\ket{\phi_{0}}+\\
&~~~~2\textup{Re}(\bra{\phi_{0}}OU(x)\ket{\phi_{0}})\big)\\
\end{split}
\label{eqfR}
\end{equation} 
It is clear that the first term produces the same frequency components as the traditional encoding scheme, whereas the second term corresponds to the zero-frequency component, independent of input feature $x$. So the key lies in the third term. Because the eigenstates $\ket{h_{j}}$ of the generator Hamiltonian form a complete basis, we can then expand the initial quantum state $\ket{\phi_{0}}$ and the observable $O$ as $\ket{\phi_{0}}=\sum_{k}\phi_{k}\ket{h_{k}}$ and $O=\sum_{j,k}o_{jk}\ket{h_{j}}\bra{h_{k}}$. By using the equation $U(x)\ket{h_{j}}=e^{iw_{j}x}\ket{h_{j}}$, we can have
\begin{equation}
\begin{split}
\bra{\phi_{0}}OU(x)\ket{\phi_{0}}
&=\sum_{j,k}\phi_{j}^{*}\bra{h_{j}}o_{jk}\ket{h_{j}}\bra{h_{k}}U(x)\phi_{k}\ket{h_{k}}\\
&=\sum_{j,k}(\phi_{j}^{*}o_{jk}\phi_{k})e^{iw_{k}x}\\
\label{wk}
\end{split}
\end{equation} 

It can be found that this part will produce new frequency components for the quantum models, which are the eigenfrequencies of generator themselves $\pm w_{k}$ for $k\in [d]$, but not the differences between them. Therefore, the new spectra of the one-layer data-encoding block with residual connection is  
\begin{equation}
\begin{split}
\Omega^{R}_{l=1}=\{w_{k}-w_{j},\pm w_{k}\vert j,k\in [d]\}
\end{split}
\end{equation}
which indicates that the frequency generation forms of the quantum neural networks with residual encoding is more diverse, and the resulting Fourier spectrum in general could also be more abundant. In this case, the toy model we exemplified above will produce new spectrum $\{0,\pm1/2,\pm1\}$, which includes more frequency components and leads to an enhanced approximation ability for the parameterized quantum circuits. 

A natural issue needs to be addressed is when will the residual encoding strategy behaves better than the traditional method. For the one-layer data-encoding block in quantum neural networks, it needs to meet the condition that there exists frequency component $w_{k}\notin\Omega$ for $k\in [d]$, which implies 
\begin{equation}
\begin{split}
\vert w_{j}-w_{l}\vert\neq \vert w_{k}\vert,~\forall~ j,l\in [d],~\exists~k\in [d]
\end{split}
\end{equation}
Such a constraint can be satisfied in many practical cases because we usually use Pauli operators as the generator Hamiltonian. 

Furthermore, for the data-encoding strategy repeated $l$-times either in sequence or in parallel, the traditional scheme will lead to a frequency spectrum $\Omega_{l}=\{(w_{j_1}+\cdots w_{j_l})-(w_{k_1}+\cdots w_{k_l})\vert j_1,\cdots,j_l,k_1,\cdots,k_l \in [d]\}$, which has only one frequency combination form, namely the difference between the sum of two sets of $l$ frequencies \cite{schuld2021effect}. However, for the residual encoding, there are more ways to construct the spectrum and the combination forms of frequencies will be more complex and diversified. Specifically, the frequency spectrum of a two-layer residual encoding is
\begin{equation}
\begin{split}
\Omega^{R}_{l=2}=\{ &(w_{j_1}+w_{j_2})-(w_{k_1}+w_{k_2}), \\
&\pm(w_{j_1}+w_{j_2}), (w_{j_1}-w_{k_1})\\
&\pm (w_{j_1}+w_{j_2}-w_{k_1}), \vert j_1,j_2,k_1,k_2 \in [d]\}
\end{split}
\end{equation}
which contains four kinds of frequency combination forms. More frequency generation forms in general can result in a larger upper limit for the spectrum size. We can summarize by induction that for a $l$-layer residual encoding scheme, the number of frequency combination forms is
\begin{equation}
\begin{split}
\mathcal{N}(\Omega^{R}_{l})=(\lceil l/2 \rceil+1)(\lfloor l/2 \rfloor+1)\propto \mathcal{O}(l^2)
\end{split}
\end{equation}
where $\lceil \cdot\rceil$ and $\lfloor \cdot \rfloor$ represent roundup and rounddown functions. This is a squared improvement over the traditional scheme and detail is shown in the appendix \ref{appendixl2}.

In addition to enlarging the accessible frequency spectrum, residual encoding can also improve the flexibility of the corresponding Fourier coefficients, both of which determine the expressivity of a quantum model. The enhancement comes from two aspects, one is due to the introduction of additional optimization degrees of freedom in the generalized residual operators $\mathcal{R}_{1,2}(x/\theta)$, and another one is due to the more diverse construction methods of frequency and the corresponding recombination of Fourier coefficients, which means that a single frequency component can be generated from the recombination of different terms in the residual loss functions. The latter one is the reason why residual operator $\mathcal{R}(x)$ can behave better than the traditional encoding strategy in expanding Fourier coefficient space without introducing additional optimization parameters. We will show the expressivity improvement in detail in the numerical simulation section.


\subsection{Measurement Scheme}

To get the expectation values of an observable $O$ for the quantum state $\mathcal{R}(x)\ket{\phi_{0}}$, which is embedded in the $\ket{0}\bra{0}$ subspace of the ancillary qubit,  we can introduce another observation operator $\bar{O}=\ket{0}\bra{0}\otimes O$ on the system. Then the output observation values can be expressed as
\begin{equation}
\begin{split}
\bar{f}_{R}(x,\theta)&=\bra{\phi_{f}}\bar{O}\ket{\phi_{f}}\\
&=\bra{0}\bra{\phi_{0}}\mathcal{R}^{\dagger}(x) (\ket{0}\bra{0}\otimes O)\ket{0}\mathcal{R}(x)\ket{\phi_{0}}\\
&=f_{R}(x,\theta)
\end{split}
\end{equation} 
where $\ket{\phi_{f}}=\ket{0}\mathcal{R}(x)\ket{\phi_{0}}+\ket{\bot}$ is the output quantum state of the whole system, and the second item $\ket{\bot}$ is orthogonal to the first part. Furthermore, because we can expand the measurement operator as $\bar{O}=(\sigma_{0}+\sigma_{z})/2\otimes O$, we can also have 
\begin{equation}
\begin{split}
\bar{f}_{R}(x,\theta)=\frac{1}{2}\big(\langle \sigma_{0}\otimes O\rangle+\langle \sigma_{z}\otimes O\rangle\big)\\
\end{split}
\end{equation} 
This indicates that we can obtain the residual loss functions $f_{R}(x,\theta)$ by measuring the average expectation of system output state $\ket{\phi_{f}}$ with two observations $\{\sigma_{0}\otimes O, \sigma_{z}\otimes O\}$, which is experimentally feasible and introduces little resource overhead. For a $l$-layer residual encoding, we need $l$ ancillary qubits at most and the corresponding observation operators will be $\{(\sigma_{0}+\sigma_{z})^{\otimes l}\otimes O\}$, whose size grows exponentially with layers of residual encoding. In practice, we do not need to use residual feature maps in every block, and inserting residual connections to some sampled data-encoding blocks could make the networks obtain better expressivity. In addition, the measurement schemes suggest that our algorithm is compatible with the existing methods for calculating the gradient of expectation value of the quantum circuit with respect to the optimization parameters \cite{schuld2019evaluating,mari2021estimating,wierichs2022general}. Using parameter-shift rule \cite{schuld2019evaluating}, the gradient of the residual loss function for a parameter $\theta_{j}$ can be calculated as
\begin{equation}
\begin{split}
\frac{\partial f_{R}(x,\theta)}{\partial \theta_{j}}=\frac{1}{2}\big[f_{R}(x,\theta_{j}+\frac{\pi}{2})-f_{R}(x,\theta_{j}-\frac{\pi}{2})\big]\\
\end{split}
\end{equation} 
where $f_{R}(x,\theta_{j}\pm\pi/2)$ are the expectation values when the target parameter $\theta_{j}$ is shifted by $\pm\pi/2$ respectively.

Furthermore, it should be mentioned that the approximation improvement can be understood from the universal approximation property with polynomial basis functions \cite{goto2021universal}, which states that the linear combination of different observations can approximate any continuous functions. Based on the above analysis for the quantum models with the specific residual encoding structures, we can see that such a combination of measurement results can actually lead to a frequency richness improvement in the Fourier series, which enhances the expressivity ability of quantum neural networks. Therefore, our work can serve as a specific case to bridge the polynomial approximation \cite{goto2021universal} and Fourier series approximation \cite{schuld2021effect}, two perspectives for understanding the universal approximation property of quantum machine learning models.

 
\section{Numerical Demonstration}\label{ChapSim}

\begin{figure}
\centering
\includegraphics[width=\linewidth]{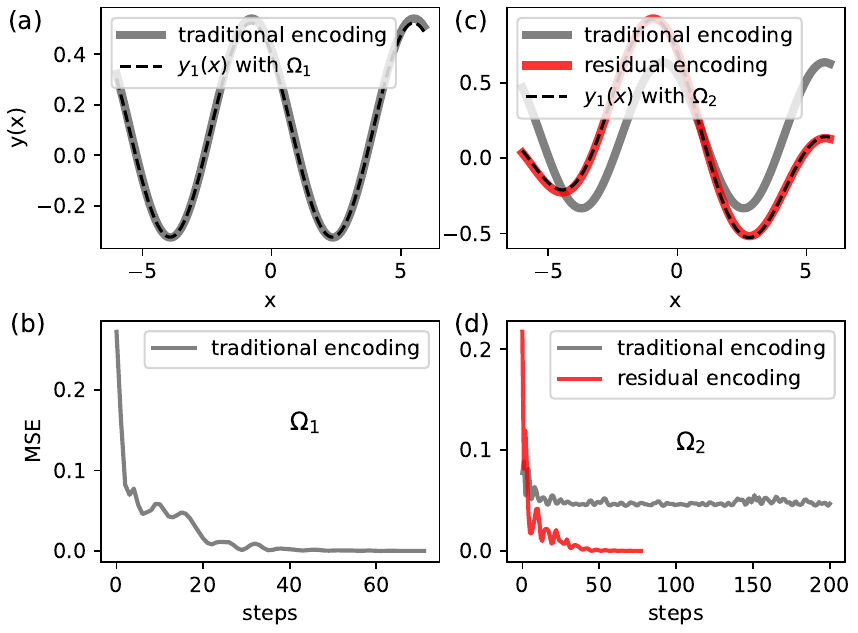}
\caption{The fitting results of quantum models to the target function $y_{1}(x)$ with frequency spectra $\Omega_{1}=\{0,1\}$ (a,b) and $\Omega_{2}=\{0,1,0.5\}$ (c,d). The top panels show the theoretical function values (black dashed lines), and the quantum model outputs with traditional (gray) and residual (red) encoding strategies, respectively. The bottom panels show the MSE values during the training processes.}
\label{fig2demo}
\end{figure}

To demonstrate the improvement of the Fourier frequency spectrum by residual connections, we present a proof-of-principle numerical simulation with Pennylane \cite{bergholm2018pennylane} here, which solves regression tasks of fitting quantum models to the target Fourier series. We adopt the traditional qubit encoding strategy to map classical data $x$ into quantum state with a single-qubit Pauli-rotation $U(x)=R_y(x)=e^{-ix\sigma_{y}/2}$ operator, where the generator Hamiltonian $G=-\sigma_{y}/2$ has two eigenvalues $e_{1,2}=\pm1/2$. The optimization ansatz used has two arbitrary single-qubit rotation gates $U(\theta_{i})=R_z(\theta_{i}^{1})R_y(\theta_{i}^{2})R_z(\theta_{i}^{3})$ for $i=1,2$ placed before and after the data-encoding block, resulting a quantum model $U_{\theta}(x)=U(\theta_{2})U(x)U(\theta_{1})$. The observable is $\sigma_{z}$ and then the loss function is $f(x,\theta)=\bra{0}U_{\theta}^{\dagger}(x)\sigma_{z}U_{\theta}(x)\ket{0}$. The quantum models are trained by a supervised learning frame to search the optimal parameters $\theta^*$, which minimizes the mean squared error (MSE) as 
\begin{equation}
\begin{split}
\Delta(\theta)=\frac{1}{2D}\sum_{i=1}^{D}(y(x_i)-f(x_{i},\theta))^2
\end{split}
\end{equation} 
where $D$ is the dimension of the data set and $y(\cdot)$ is the target function. We use Adam optimizer with at most 200 steps and set the learning rate as 0.3 with batch size $0.7D$ in the simulation. A termination condition for the optimization convergence, that is, the variance of ten consecutive loss function values is less than $10^{-8}$, is also used. 

\begin{figure}
\centering
\includegraphics[width=\linewidth]{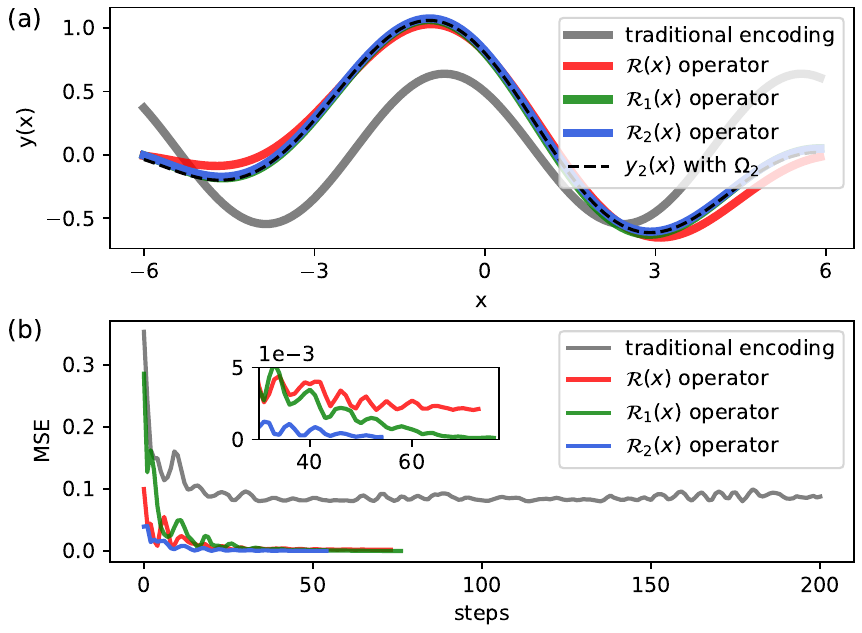}
\caption{(a) The fitting results of quantum models to the target function $y_{2}(x)$ with traditional encoding scheme (gray) and residual feature map with the $\mathcal{R}(x)$ (red), $\mathcal{R}_{1}(x)$ (green) and $\mathcal{R}_{2}(x)$ (blue) operators, respectively. (b) The MSE values during the training processes.}
\label{figdemo2}
\end{figure}

As shown in the figure \ref{fig2demo}, this quantum model can learn functions of the form $y_{1}(x)=\sum_{\omega_i\in\Omega_{1}}(ae^{i\omega_{i}x}+a^*e^{-i\omega_{i}x})$ with a MSE value $\Delta=6.0\times10^{-5}$, where $a$ is an amplitude parameter and the frequency spectrum is $\Omega_{1}=\{\omega_0=0,\omega_1=2\vert e_{1,2}\vert=1\}$, and this is consistent to the results in \cite{schuld2021effect}. However, a multi-frequency function with spectrum $\Omega_{2}=\{\omega_0=0,\omega_1=1,\omega_2=0.5\}$ cannot be well fitted with error $\Delta=5.1\times10^{-2}$, due to the frequency lack of parameterized quantum circuits caused by data-encoding strategy. The frequency mismatch can be mitigated by inserting residual connections to the data-encoding block with an output MSE value $\Delta=5.1\times10^{-5}$, because the resulting residual operator $\mathcal{R}(x)$ can bring richer frequency components to enhance the circuit expressivity. It is worth noting that the residual data encoding scheme still works well for the spectral $\Omega_{1}$ besides $\Omega_{2}$, and the optimization process can converge quickly.

\begin{figure}
\centering
\includegraphics[width=\linewidth]{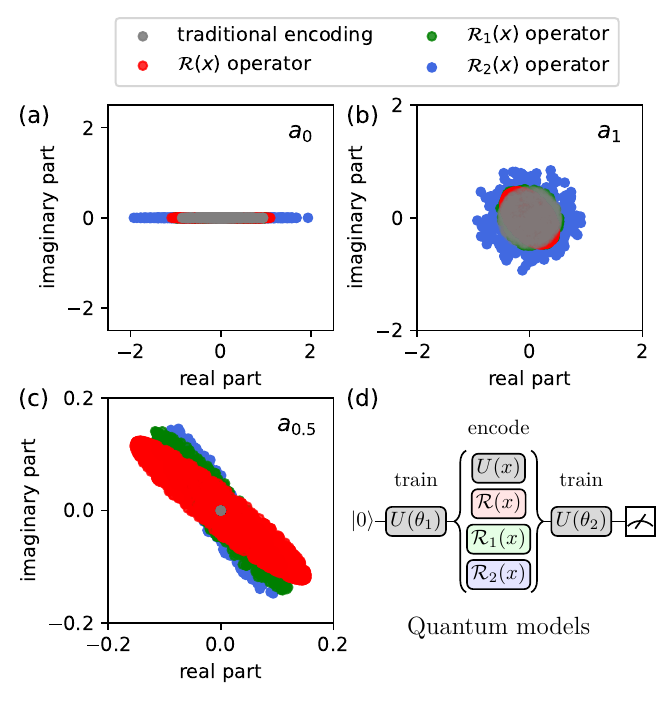}
\caption{(a-c) The real and imaginary parts of the Fourier coefficients sampled from 1000 random quantum models. (d) Quantum models with one-layer data-encoding structure. The quantum models share the same ansatz but vary the data-encoding strategies by traditional encoding (gray), residual feature map with the $\mathcal{R}(x)$ (red), $\mathcal{R}_{1}(x)$ (green) and $\mathcal{R}_{2}(x)$ (blue) operators. The distribution of coefficients widens from gray to red to green to blue.  }
\label{figcoff}
\end{figure}

Furthermore, we turn to a more general case for fitting the function $y_{2}(x)=\sum_{\omega_i\in\Omega_{2}}(a_{\omega_i}e^{i\omega_{i}x}+a_{\omega_i}^*e^{-i\omega_{i}x})$, where the amplitudes can be different for each frequency component. Additional degrees of freedom can be obtained from the multi-combination methods of single-frequency components in residual loss functions and the parameterized gates on the auxiliary qubit in the generalized residual operators $\mathcal{R}_{1,2}(x/\theta)$. We can conclude from the numerical results in the figure \ref{figdemo2} that the traditional encoding scheme still cannot fit the target function with MSE value $\Delta=0.09$, while the residual feature map with $\mathcal{R}(x)$ operator works better with error $\Delta=2.1\times10^{-3}$. When we use the generalized residual operators, the fitting results can be further improved, which converges to a smaller MSE values with $\Delta=1.1\times10^{-4}$ for $\mathcal{R}_{1}(x)$ and $\Delta=1.7\times10^{-4}$ for $\mathcal{R}_{2}(x)$ in fewer optimization steps with $77$ steps for $\mathcal{R}_{1}(x)$ and $55$ steps for $\mathcal{R}_{2}(x)$. Moreover, the extra combination forms and trainable parameterized quantum gates bring more flexibility for fitting, which expand the Fourier coefficient space. As shown in the figure \ref{figcoff}, we sample the quantum models 1000 times with different feature maps which produce Fourier series, and then get the distribution of Fourier coefficients. We can see that under the same ansatz, the residual feature map with $\mathcal{R}_{2}(x)$ operator has the widest Fourier coefficients distribution, and all the three residual encoding are better than the traditional encoding scheme. 

\begin{figure}
\centering
\includegraphics[width=\linewidth]{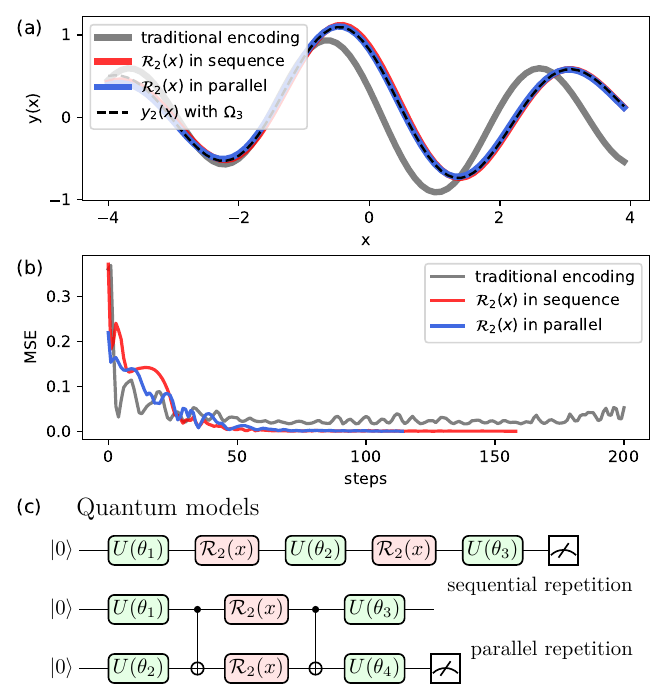}
\caption{(a) The fitting results of quantum models with two-layer data-encoding for target function $y_{2}(x)$ with frequency spectra $\Omega_{3}$. (b) The MSE values during the training processes. (c) Quantum models with two-layer data-encoding structure. The residual operator $\mathcal{R}_{2}(x)$ is repeated in sequence and in parallel, and the output is the measurement value $\langle\sigma_{z}\rangle$ on a qubit. }
\label{fig2layer}
\end{figure}

In addition, this enhancement can be quantitatively measured by a commonly used expressibility metric \cite{sim2019expressibility}. We first generate many pairs of parameters $\Theta_{1}$ and $\Theta_{2}$ randomly, and calculate the distribution ($P_F$) of state fidelities $F=\vert\bra{0}U_{\Theta_{1}}^{\dagger}(x)U_{\Theta_{2}}(x)\ket{0}\vert^2$, which measure the overlap of quantum states generated by quantum models. Then the Kullback-Leibler (KL) divergence \cite{kullback1951information} is used to quantify the circuit expressivity by comparing the sampled fidelity distributions with that of the Haar-distributed state ensemble ($P_{\textup{Haar}}$) as
\begin{equation}
\begin{split}
D_{KL}(P_F||P_{\textup{Haar}})=\sum_{j}P_F(j)\log\frac{P_F(j)}{P_{\textup{Haar}}(j)}
\end{split}
\end{equation} 
where the analytical form of the fidelity distribution for the ensemble of Haar random states is $p_{\textup{Haar}}(F)=(N-1)(1-F)^{N-2}$ and $N$ is the dimension of Hilbert space \cite{zyczkowski2005average}. A smaller KL divergence value corresponds to a more favorable expressibility. We sample each quantum model in the figure \ref{figcoff} by 1000 times and use 45 histogram bins to estimate the fidelity distribution, which are then compared with the sampled fidelities ensemble of the Haar random states. The computed results of KL divergence are $D_{KL}^{\textup{trad}}=0.0634$, $D_{KL}^{\mathcal{R}(x)}=0.0581$, $D_{KL}^{\mathcal{R}_{1}(x)}=0.0446$ and $D_{KL}^{\mathcal{R}_{2}(x)}=0.0429$, respectively. We can see that the generalized residual operators can indeed increase the circuit expressivity relative to traditional encoding scheme. Moreover, it worth mentioning that the reasons for expressivity enhancement are different for $\mathcal{R}(x)$ and $\mathcal{R}_{1,2}(x)$ operators. The former one is due to the diverse construction methods of frequencies in residual loss function, while the latter is also due to the additional optimization parameters. It is known that constructing frequencies only from the difference between the sum of the generator's eigenvalues will limit the access to higher-order components, resulting in a reduction in coefficient variance \cite{schuld2021effect}. Therefore, the residual encoding method which can offer more methods to construct frequency could broaden the distribution of Fourier coeﬀicients, which suggests an enhanced expressivity of quantum models by residual connections. 

Moreover, similar to the traditional encoding, we can extend the accessible frequency spectrum by repeating the residual encoding block multi-times in sequence or in parallel method. To investigate the frequency extension by sequential and parallel repetitions of data-encoding, we fit the aforementioned target function $y_{2}(x)$ with a more complex spectra $\Omega_{3}=\{\omega_0=0,\omega_1=1,\omega_2=0.5,\omega_3=1.5,\omega_4=2\}$ and amplitude $a_{0}=0.1$ and $a_{1.5,2}=5a_{1,0.5}=0.15+0.15i$. Two-layers of repeating structures for the traditional encoding in sequence and residual encoding with $\mathcal{R}_2(x)$ operators in sequence and in parallel are used, as shown in the figure \ref{fig2layer}. The single-qubit observable is $O=\sigma_{z}$ for all cases. All the quantum models were trained with 200 steps at most using Adam optimizer and with batch size 16. We can see that both the sequential and parallel repetitions of residual encoding can extend the Fourier spectrum and fit the target function well. The MSE values and optimization steps for the sequential repetitions are $\Delta=3.3\times10^{-4}$ and $159$ steps, while $\Delta=4.2\times10^{-4}$ and $115$ steps for parallel repetitions. It should be clarified that the mixed use of residual and traditional encoding will also bring an enhanced expressivity. Therefore, replacing parts of the encoding blocks in complex quantum models with residual blocks, but not all of them, can enrich the expressivity of the whole neural networks.


\section{Application in image classification}\label{ChapExp}

In this part, we turn to discuss the performance of QCNN algorithm with residual encoding for image classification using a real-word dataset MNIST. The MNIST includes 60000 (10000) images for train (test) datasets with 10 classes of handwritten digits, and each image is a $28\times28$ pixels data. Here we focus on the binary classification with selected classes 0 and 1, and the sizes for the train and test datasets used are 12665 and 2115. Constrained by the current quantum hardwares, high-dimensional data usually require classical pre-processing techniques for dimensionality reduction, and we adopt principal component analysis (PCA) technology to match the input data with the four-qubit data-encoding layer \cite{jolliffe2016principal}. For comparison, we use qubit encoding and consider the case where no residual connection is added, and the case where the residual operator $\mathcal{R}_{2}(x)$ is applied to the $i$-th qubit, denoted as traditional and residual-$Q_i$ schemes, respectively

 \begin{figure}
\centering
\includegraphics[width=\linewidth]{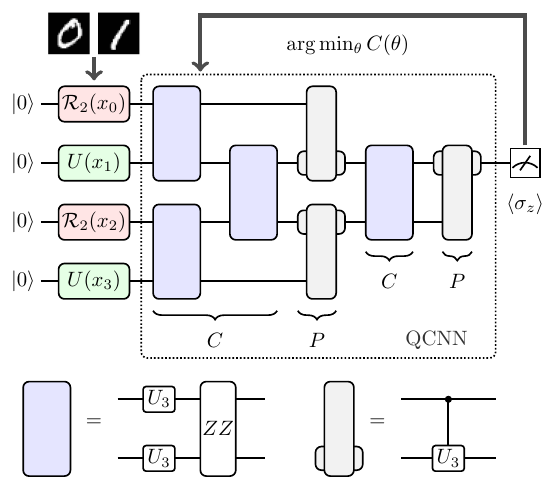}
\caption{A schematic of the QCNN algorithm with residual encoding for image classification. The handwritten digits are encoded as quantum states via quantum feature map, where the green blocks represent qubit encoding schemes and the red blocks are residual encoding with $\mathcal{R}_{2}(x_{i})$ operators on the $i$-th qubit. The multiple convolutional ($C$) and pooling ($P$) layers use quantum gates with trainable parameters $\theta$, and the detailed structures are shown below. The measurement outcome of the quantum circuit $\langle\sigma_{z}\rangle$ is used to calculate the cost function $C(\theta)$ and characterize the binary classification results $c_{0/1}$. The classical computer updates the optimization parameters of QCNN algorithm based on gradients until the cost function converges.}
\label{figqcnn}
\end{figure}

The ansatz for QCNN algorithm is composed of a series of alternating convolutional and pooling layers \cite{cong2019quantum}, as shown in the figure \ref{figqcnn}. Each convolutional layer includes several single- and two-qubit parameterized quantum gates, keeping a translationally invariant structure. We use Ising interactions between adjacent qubits with one parameter as $ZZ(\phi)=e^{-i\sigma_{z}\otimes\sigma_{z}\phi/2}$ and single-qubit $U_3$ gates with three parameters as 
\begin{equation}
\begin{split}
&U_{3}(\theta,\phi,\delta)=\begin{bmatrix}
\cos(\theta/2)&-e^{i\delta}\sin(\theta/2)\\
e^{i\phi}\sin(\theta/2)&e^{i(\phi+\delta)}\cos(\theta/2)\\
 \end{bmatrix}
\end{split}
\end{equation} 
The pooling layer is implemented by a parameterized controlled-$U_3$ gate and one qubit will be traced out, reducing the quantum states from two qubits to a single qubit. We measure the expectation values $\langle\sigma_{z}\rangle_i$ on the output qubit for the $i$-th input data with label $y_{i}=0/1$. The cost function is $C(\theta)=\sum_{i=1}^{D}(\vert\langle\sigma_{z}\rangle\vert_{i}-y_{i})^2/2D$ for a $D$-dimensional dataset and it is optimized by Adam optimizer with a learning rate 0.2. The number of iterations in the training process is 100 and the processes are repeated 20 times to obtain the mean values with random initialization of optimization parameters. Once the cost function converges and the optimal parameters $\theta^*=\textup{arg}\min_{\theta} C(\theta)$ are obtained, the measurement outputs can be reconstructed into binary values $c_{0/1}$ via a boundary precision $\epsilon\in(0,0.5]$. We suppose that the classification result is $c_{0/1}=1$ for $\vert\langle\sigma_{z}\rangle\vert>1-\epsilon$ and $c_{0/1}=0$ for $\vert\langle\sigma_{z}\rangle\vert<\epsilon$, while other values are marked as unclassifiable optimization results. A smaller value for $\epsilon$ represents higher optimization accuracy and higher classification standards.

\begin{figure}
\centering
\includegraphics[width=\linewidth]{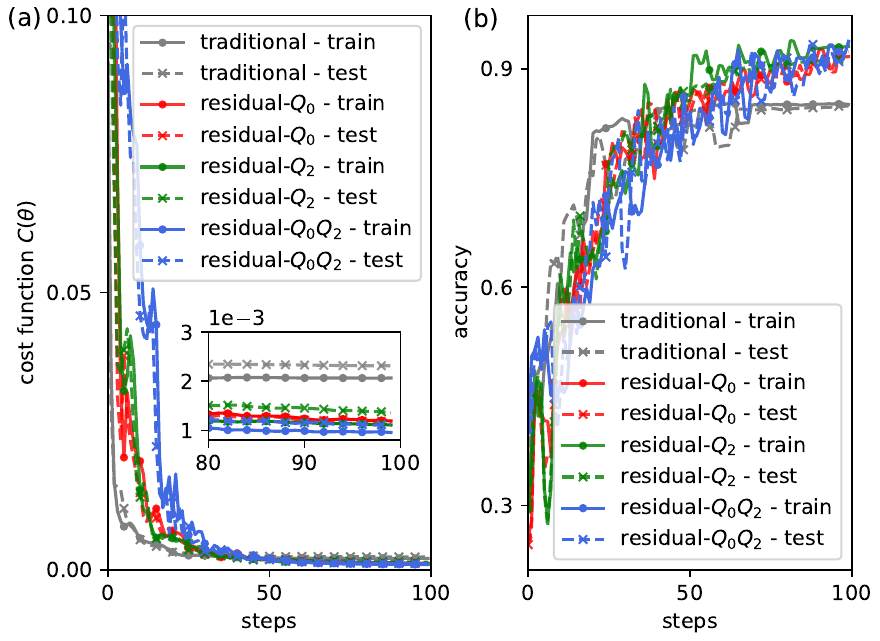}
\caption{The performance of QCNN algorithm with different data-encoding strategies for image classification. Simulations with the traditional scheme and residual encoding on qubits $Q_0$ and $Q_2$ in the train and test datasets are offered. The panel (a) shows the evolution processes of cost function with optimization steps and panel (b) is the corresponding results in accuracy.}
\label{figqcnn-result}
\end{figure}

\begin{table}
\centering
\begin{tabular}{ccccc}
\hline
\hline
dataset&traditional &residual-$Q_0$&residual-$Q_2$&residual-$Q_0Q_2$\\
\hline
train&85.10\%&91.66\%&93.10\%&93.78\%\\
test&84.90\%&92.35\%&91.41\%&93.65\%\\
\hline
\hline
\end{tabular}
\caption{The average accuracy obtained from twenty repetitions of training for the image binary classification with MNIST datasets using different data-encoding strategies.}
\label{tab-accuracy}
\end{table}

The optimization results of cost function and accuracy are shown in the figure \ref{figqcnn-result} and table \ref{tab-accuracy}. We set $\epsilon=0.1$ in the simulation and there are 20 free parameters involved in the ansatz. We can conclude that the residual encoding schemes can obtain smaller convergence values of loss than the traditional encoding method, which means that the models have better approximation ability. Such an enhancement can lead to better expressivity and higher accuracy for quantum models in complex learning tasks. In addition, the residual encoding can produce a high classification accuracy, reaching 92.85\% and 92.47\% on average for the train and test datasets respectively, which are about 7.74\% and 7.57\% higher than that with the traditional encoding strategy.

 
\section{Conclusion}\label{ChapCon}

In summary, we have proposed a complete quantum circuit-based architecture for the digital implementation of quantum residual neural networks, dubbed QResNets. The classical residual connection channel is quantized by adding an auxiliary qubit to the data-encoding and trainable blocks, which is then generalized with additional parameterized gates. We further prove mathematically that the Fourier spectrum of quantum models output can be enriched when the residual connections are applied to the data-encoding blocks. There is a squared improvement in the number of frequency generation forms of residual encoding over the traditional schemes. It means that the $l$-layer residual encoding strategy can produce $\mathcal{O}(l^2)$ frequency combination methods, rather than just by the difference of sum of generator eigenvalues as in traditional methods. Moreover, the diverse spectrum construction methods in the residual loss functions and additional optimization degrees of freedom in the generalized residual operators could make the Fourier coeﬀicients more flexible, favoring the access to higher-order components. This indicates that the residual encoding can enrich the spectrum and broaden the Fourier coefficient distribution, that is, it can enhance the expressivity of various parameterized quantum circuits. Various numerical simulation of fitting the functions of Fourier series, and a demonstration of binary classification in images of handwritten digits with MNIST datasets are conducted to show the algorithm performance. Compared with the traditional encoding, the accuracy of residual encoding can be improved by about seven percent. Our work advances the design of quantum neural networks with specific structures and, for the first time, enables a full quantum realization of classical residual connections, and also provides a new quantum feature map strategy.


\section{Acknowledgements} 

We acknowledge the support from the National Key R\&D Plan (2021YFB2801800).

\appendix

\section{Generalized Residual Operators }\label{appendixR2}

We have discussed the form of residual operator $\mathcal{R}(x/\theta)$ and its corresponding residual loss function $f_{R}(x,\theta)$ above. In this part, we give a detail introduction to the generalized residual operators $\mathcal{R}_{1,2}(x)$ and the corresponding generalized residual loss function $f_{R_{1,2}}(x,\theta)$, which present stronger expressivity. As shown in equation \ref{R1op} where one Hadamard gate is replaced by a parameterized gate, we further assume that both two Hadamard gates on the ancillary qubit are replaced by gates $R_{y}(2\alpha)$ and $R_{y}(2\gamma)$ with trainable angles $\alpha$ and $\gamma$, then the $\mathcal{R}_{2}(x/\theta)$ operator can be expressed as 
\begin{equation}
\begin{split}
\mathcal{R}_{2}(x/\theta)=\cos\alpha\cos\eta \sigma_{0}^{\otimes n}+\sin\alpha\sin\eta\cdot\mathcal{L}(x/\theta)
\end{split}
\end{equation} 
with a relabeled angle $\eta=\pi m_a/2-\gamma$. The residual operator $\mathcal{R}_{1}(x/\theta)$ can be seen as a special case with $\gamma=-\pi/4$ ignoring a global phase factor. When the generalized residual operator $\mathcal{R}_{1,2}(x)$ is used in the data-encoding block, the residual loss function is
\begin{equation}
\begin{split}
f_{R_{1,2}}(x,\theta)=&\bra{\phi_{0}}\mathcal{R}_{1,2}^{\dagger}(x)O\mathcal{R}_{1,2}(x)\ket{\phi_{0}}\\
=&A_1^{R_{1,2}}f(x,\theta)+A_2^{R_{1,2}}\bra{\phi_{0}}O\ket{\phi_{0}}+\\
&A_3^{R_{1,2}}\textup{Re}(\bra{\phi_{0}}OU(x)\ket{\phi_{0}})\\
\end{split}
\label{eqfR2}
\end{equation} 
where the trainable coeﬀicients for $\mathcal{R}_{1}(x)$ operator are $A_1^{R_{1}}(\alpha)=\sin^2\alpha/2$, $A_2^{R_{1}}(\alpha)=\cos^2\alpha/2$ and $A_3^{R_{1}}(\alpha)=(-1)^{m_a}\sin2\alpha/2$, while for the $\mathcal{R}_{2}(x)$ operator are $A_1^{R_{2}}(\alpha,\eta)=(\sin\alpha\sin\eta)^2$, $A_2^{R_{2}}(\alpha,\eta)=(\cos\alpha\cos\eta)^2$ and $A_3^{R_{2}}(\alpha,\eta)=(\sin2\alpha\sin2\eta)/2$. Such extension offers additional degree of freedom for the optimization process and can relax the range of Fourier coefficients for the new frequency component $w_k$ in equation \ref{wk} to $A_3^{R_{1,2}}\sum_{j}\phi_{j}^{*}o_{jk}\phi_{k}$, and similar effect is true for other frequency components. In fact, the generalized residual loss function $f_{R_{1,2}}(x,\theta)$ can be seen as a weighted version of the residual loss function $f_{R}(x,\theta)$, where the weights of each term are trainable.


\section{Proof of Frequency Combination Forms}\label{appendixl2}

As mentioned above, there are four kinds of combination forms for frequency generation with a two-layer residual encoding. When another residual encoding layer is added, the spectrum $\Omega^{R}_{l=1}=\{w_{k}-w_{j},\pm w_{k}\vert j,k\in [d]\}$ would be combined to the spectrum $\Omega^{R}_{l=2}$. We first consider the component of difference of the sum of generator eigenvalues, and it would bring new frequency components for the three-layer residual spectrum as
\begin{equation}
\begin{split}
\big\{ 
&\sum_{m=1}^{3}w_{j_m}-\sum_{n=1}^{3}w_{k_n},\pm(\sum_{m=1}^{3}w_{j_m}-\sum_{n=1}^{2}w_{k_n})\\
&\pm(\sum_{m=1}^{3}w_{j_m}-w_{k_1}),\sum_{m=1}^{2}w_{j_m}-\sum_{n=1}^{2}w_{k_n}
\big\}
\end{split}
\label{B1}
\end{equation}
with index $ j_1,j_2,j_3,k_1,k_2,k_3 \in [d]$. If we further consider the effect of eigenvalues $\pm w_{k}\in\Omega^{R}_{l=1}$, more frequency components can be involved as 
\begin{equation}
\begin{split}
\big\{ 
&\pm(\sum_{m=1}^{3}w_{j_m}-\sum_{n=1}^{2}w_{k_n}),\pm(\sum_{m=1}^{3}w_{j_m}-w_{k_1}),\pm\sum_{m=1}^{3}w_{j_m},\\
&\pm(\sum_{m=1}^{2}w_{j_m}-\sum_{n=1}^{2}w_{k_n}),\pm(\sum_{m=1}^{2}w_{j_m}-w_{k_1})
\big\}
\end{split}
\label{B2}
\end{equation}

We can combine the above cases for frequency generation and simply mark the combination forms of $\pm(\sum_{m=1}^{l_1\ge1}w_{j_m}-\sum_{n=1}^{l_2\ge1}w_{k_n})$ as $\ds{l_1}{l_2}$, which means the difference between the sum of two sets with $l_1$ and $l_2$ frequencies. Note that we mark the combination form of $\pm\sum_{m=1}^{l\ge1}w_{j_m}$ as $\ds{l}{0}$. Then we can find that there are six kinds of frequency combination forms for the three-layer residual encoding, and it can be concluded as $\{\ds{3}{3},\ds{3}{2},\ds{3}{1},\ds{3}{0},\ds{2}{2},\ds{2}{1}\}$. Further, for the $l$-layer residual encoding, the spectrum with various frequency generation forms can be formally expressed as
\begin{equation}
\begin{split}
\Omega^{R}_{l}=\big\{
&\ds{l}{l},\ds{l}{l-1},\cdots,\ds{l}{1},\ds{l}{0}\\
&\ds{l-1}{l-1},\cdots,\ds{l-1}{1}\\
& \cdots \\
& \ds{\lceil l/2\rceil}{\lfloor l/2\rfloor}
\big\}
\end{split}
\label{B3}
\end{equation}
where the $\lceil \cdot\rceil$ and $\lfloor \cdot \rfloor$ are roundup and rounddown functions. Based on the number of items in each row of equation \ref{B3}, we can determine the number of components in the set as
\begin{equation}
\begin{split}
\mathcal{N}&(\Omega^{R}_{l})=(l+1)+(l-1)+\cdots+(\lceil l/2\rceil-\lfloor l/2\rfloor+1)\\
&=\frac{(l+2)+(\lceil l/2\rceil-\lfloor l/2\rfloor)}{2}\frac{(l+2)-(\lceil l/2\rceil-\lfloor l/2\rfloor)}{2}\\
&=(\lceil l/2 \rceil+1)(\lfloor l/2 \rfloor+1)
\end{split}
\end{equation}

It can be concluded that compared with the traditional encoding method which generates frequency only with $\ds{l}{l}$ \cite{schuld2021effect}, there is a squared improvement in frequency generation methods for the residual encoding scheme with $\mathcal{N}(\Omega^{R}_{l})\propto\mathcal{O}(l^2)$. While different combinations may produce some of the same frequency components, in general, more frequency-generation methods suggest that the possible upper bounds for the size of Fourier spectrum of quantum model outputs can be larger, allowing for more complex learning tasks. Moreover, the diverse construction methods for frequencies can also improve the flexibility of Fourier coefficients, favoring the access to higher-order components and further improving the expressivity of quantum models.

 
\bibliographystyle{unsrtnat}
\bibliography{QResNets_wen.bib}

 
\end{document}